\documentclass[aps,prd,preprint,preprintnumbers,nofootinbib]{revtex4-1}

\usepackage{amsmath}
\usepackage{graphicx}
\usepackage{epstopdf}
\usepackage{amsfonts,amssymb}
\usepackage{float}
\usepackage[utf8]{inputenc}
\usepackage{pstricks}
\usepackage{color}
\usepackage[compat=1.0.0]{tikz-feynman}
\usepackage{simplewick}
\usepackage{color, colortbl}
\definecolor{Gray}{gray}{0.9}
\usepackage{float}
\usepackage{tikz-feynman}
\usepackage{feynmp}
\usepackage{forest}

\usepackage{verbatim}    
\usepackage{dcolumn}
\usepackage{bm}
\usepackage{epsfig}
\usepackage{braket}
\usepackage[normalem]{ulem} 
\usepackage{longtable,booktabs}
\usepackage{booktabs}

\usepackage{multirow}
\usepackage{verbatim}
\usepackage{hyperref}
\usepackage{tabularx}
\usepackage{bm}
\usepackage{epsfig}
\usepackage{braket}
\usepackage{longtable,booktabs}
\usepackage{tikz}
\usetikzlibrary{tikzmark}
\usepackage[normalem]{ulem}

\newcommand{\be}{\begin{equation}}
\newcommand{\ee}{\end{equation}}
\newcommand{\ben}{\begin{eqnarray}}
\newcommand{\een}{\end{eqnarray}}

\begin{document}

\title{Searching for femtoscopic signatures of the $D \bar D (I=0) \ [X(3700)]$ bound state}
\author{Luciano M. Abreu}
\email{luciano.abreu@ufba.br}
\affiliation{ Instituto de F\'isica, Universidade Federal da Bahia,
Campus Universit\'ario de Ondina, 40170-115, Bahia, Brazil}

\author{Juan M. Torres-Rincon}
\email{torres@fqa.ub.edu}
\affiliation{Departament de F\'isica Qu\`antica i Astrof\'isica and Institut de Ci\`encies del Cosmos (ICCUB), Facultat de F\'isica, Universitat de Barcelona, Mart\'i i Franqu\`es 1, 08028 Barcelona, Spain}

\date{\today}

\begin{abstract}

The femtoscopic $D \bar D $ correlation functions are investigated to predict the signature of the not-yet-established $X(3700)$ state. Here it is interpreted as a bound state generated by solving the coupled-channel Bethe-Salpeter equations with the local hidden-gauge formalism. We prospect how the relevant properties and observables characterizing this state---like the pole position, scattering lengths and compositeness---might be affected by the variation of the parameters of the model. The amplitudes are then used as input into the momentum correlation functions of the $D^0 \bar D^0$ and  $D^+ D^- $ pairs. We discuss how their behaviors encode the features of the $X(3700)$ state. 

\end{abstract}

\maketitle

\section{Introduction}

Among the new hadrons observed in the last decades~\cite{ParticleDataGroup:2024cfk}, there are several states with minimum valence quark content $c \bar{c} q \bar{q}$ and properties that are incompatible with the quark model predictions~\cite{Brambilla:2019esw,Chen:2022asf,Meng:2022ozq}. As a consequence, the study of these exotic hadron states has emerged as one of the most fascinating topics in the contemporary hadron physics.  
These multiquark structures have been interpreted by distinct configurations in literature, for instance: weakly bound meson molecules, compact tetraquark states, excited conventional hadrons, cusps engendered by kinematical singularities, glueballs, hybrids, etc., or even a superposition of some of them. 
A classic example of this complex landscape is the famous state $X(3872)$ (or $\chi_{c1}(3872)$), first observed in 2003 by the Belle Collaboration~\cite{Belle:2003nnu} and afterwards, confirmed by other experiments~\cite{ParticleDataGroup:2024cfk}. Its quantum numbers have been established to be $I^G(J^{PC}) = 0^+(1^{++})$, 
and one of the most used configurations for its intrinsic nature is the weakly bound state of open charmed mesons $(D\bar D^{*} +c.c.)$, due to its proximity to the $D\bar D^{*}$ threshold~\cite{Brambilla:2019esw,Chen:2022asf,Meng:2022ozq} .
In particular, many works using effective-field theories, lattice-QCD calculations~\cite{Prelovsek:2013cra}, Dyson-Schwinger equations~\cite{Wallbott:2019dng} and other approaches claimed for the prominent molecular component of this state. Notwithstanding, it has also been stated that a $c \bar c$ component can be sizable; see for example the lattice-QCD result in Ref.~\cite{Prelovsek:2013cra}, and the theory works~\cite{Song:2023pdq,Terashima:2023tun}. Besides, one can also state that other authors claimed that it could be a $c \bar c q \bar q $  compact tetraquark state. See, for example, the recent review~\cite{Esposito:2025hlp}.
So, the internal structure  of the $X(3872)$ remains a matter of debate. Similar disputes also happen with other observed structures.
So far, there is no conclusive and universal understanding of the exotic collection. To address this issue and set up distinctions among the different interpretations, observables like the masses, decay widths and production rates have been studied theoretically and experimentally. 

In parallel with the observation of the $X(3872)$ and the controversy about its interpretation, a natural consequence in the scenario of meson molecule configuration was the speculation on the existence of its lightest partner: the $D\bar{D}$ state, which is also commonly denoted as $X(3700)$ or $X(3720)$, with $ I^G(J^{PC}) = 0^+(0^{++})$ (henceafter, it will be denoted as $X(3700)$). 
Initially predicted using a coupled channel unitary approach~\cite{Gamermann:2006nm},
it has been theoretically investigated in different contexts, for example: in the $D \bar{D}$ mass distribution of $e^+ e^- \to J/\psi D \bar{D}$ reaction~\cite{Gamermann:2007mu,Wang:2019evy}; in the analysis of meson-meson interactions within the heavy meson effective theory~\cite{Nieves:2012tt,Hidalgo-Duque:2012rqv,Ding:2020dio}; in the $\eta \eta^{(\prime)}$ invariant mass of the radiative decays of $\psi (3770), \psi (4040)$ and of the $e^+ e^- \to J/\psi \eta \eta^{\prime}$ and  $B^+ \to K^+ \eta \eta $ reactions~\cite{Xiao:2012iq,Brandao:2023vyg}; in the $D^0 \bar{D}^0$ mass distribution of the $\psi (3770) \rightarrow \gamma D^0 {\bar{D}}^0$ decay~\cite{Dai:2020yfu}; in the coupled $D \bar{D}, D_s \bar{D}_s$ scattering on lattice~\cite{Prelovsek:2020eiw}; in $\gamma \gamma \to D \bar{D}$ reaction~\cite{Wang:2020elp,Deineka:2021aeu}; in $B$ decays~\cite{Xie:2022lyw}; in $\gamma \gamma \to D^+ D^-$ reaction in ultra-peripheral heavy ion collisions~\cite{Sobrinho:2024tre}; and others. On the other hand, experimental searches reported in the literature are scarce. We are aware of the studies made by Belle and BaBar Collaborations, which analyzed respectively the $e^+ e^- \to J/\psi D \bar{D}$ and  $e^+ e^- \to  D \bar{D}$ reactions~\cite{Belle:2007woe,Belle:2017egg,BaBar:2010jfn}. Some authors claim that these data support the $X(3700)$ observation~\cite{Gamermann:2007mu,Xiao:2012iq,Wang:2020elp,Deineka:2021aeu}. However, it is a  subject of controversy, and this state has not been yet listed in the summary tables of the Review of Particle Physics~\cite{ParticleDataGroup:2024cfk}.

In that regard, the assessment of femtoscopic correlations appears as a promising scenario. Although the idea is not new~\cite{Lisa:2005dd}, recently it has emerged as a valuable tool to describe the hadron interactions; see for example recent experimental studies~\cite{STAR:2014dcy,ALICE:2018ysd,ALICE:2022yyh,ALICE:2021cpv,ALICE:2022enj,ALICE:2022uso,ALICE:2022mxo,ALICE:2023wjz,ALICE:2024bhk} and also the review~\cite{Fabbietti:2020bfg}. 
In particular, it has been employed in the study of the near-threshold resonances, since the femtoscopic two-particle momentum correlation function (CF) in high-energy collisions is sensitive to the low-energy hadron interactions (see e.g. \cite{Chizzali:2022pjd,Liu:2022nec,Feijoo:2024bvn,Abreu:2024qqo}). More recently, the ALICE Collaboration has inaugurated the era of measurements with the femtoscopy technique in the charm sector~\cite{ALICE:2022enj,ALICE:2024bhk}. On theoretical grounds, there are also investigations that propose the use of the CFs for the understanding of the hadron interactions in the charm sector~\cite{Albaladejo:2023pzq,Torres-Rincon:2023qll,Khemchandani:2023xup} and exotic hadron states~\cite{Kamiya:2022thy,Vidana:2023olz,Albaladejo:2023wmv,Feijoo:2023sfe,Liu:2023wfo,Liu:2023uly,Liu:2024nac,Liu:2024uxn}. Specifically, Ref.~\cite{Kamiya:2022thy} has suggested the analysis of the source size 
dependence of the correlation functions of the $D^0 \bar D^{*0}$ and $D^+ D^{*-}$ pairs to discriminate the nature of the $X(3872)$ as a $(D\bar D^{*} +c.c.)$ bound state.

Thus, motivated by the scenario discussed above, in this work we propose an alternative way for searching the signature of the $X(3700)$ state, via the femtoscopic $D \bar D $ correlations. Assuming it as a bound state dynamically generated with the local hidden-gauge and Bethe-Salpeter formalisms, we explore how its pole position, scattering lengths and compositeness are influenced by the change of the parameters of the model. The amplitudes are then used as input into the CFs of the $D^0 \bar D^0$ and  $D^+ D^- $ pairs. Finally, we analyze how the CFs can contribute towards a better knowledge of the $X(3700)$. 

This paper is organized as follows. We start with a brief outline of the coupled-channel approach used to determine meson-meson scattering amplitudes, and analysis of the dependence of the properties of the $X(3700)$ with the parameters of our framework. Next, the  $D^0 \bar D^0$ and  $D^+ D^- $ CFs are introduced, and the results are presented and examined. Finally, we conclude with a summary and discussion of the main findings of this work.


\section{Scattering amplitudes}


We begin by presenting the formalism employed to describe the $D \bar D$  $(D_{s} \bar D_{s})$ dynamics. It has been introduced in Ref.~\cite{Gamermann:2006nm} and explored in Refs.~\cite{Gamermann:2007mu,Xiao:2012iq,Wang:2019evy,Wang:2020elp,Dai:2020yfu,Brandao:2023vyg,Sobrinho:2024tre}. Here we adopt its version described in Refs.~\cite{Bayar:2022dqa,Abreu:2023rye}, which is based on the extension of the local hidden gauge approach~\cite{Gamermann:2006nm,Bando:1987br,Harada:2003jx,Meissner:1987ge,Nagahiro:2008cv}. Accordingly, the interactions between vector and pseudoscalar mesons are treated under broken $SU(4)$ symmetry. More concretely, to describe the lowest-order amplitude of the $D \bar D (D_{s} \bar D_{s}) \to D \bar D (D_{s} \bar D_{s})$ reaction due to one-vector exchange, 
the following structure is used~\cite{Bayar:2022dqa,Abreu:2023rye}, 
\begin{align}\label{LPPV}
    \mathcal{L}_{PPV}= -ig \langle [P,\partial_\mu P] V^\mu\rangle,
\end{align}
where $ g $ is the coupling constant $g=m_V/(2f_\pi)$ ($m_V=800$ MeV, $f_\pi=93$ MeV); $ \left\langle\ \cdots \right\rangle $ denotes the trace over the flavor space; and $P,V$ stand for the  $ q \bar{q} $ matrices in the $ SU(4) $ flavor space, written in terms of pseudoscalar and vector mesons:
 \begin{align} \label{pv}
 P&=\left(\begin{array}{cccc}\frac{\eta}{\sqrt{3}}+\frac{\eta^\prime}{\sqrt{6}}+\frac{\pi^0}{\sqrt{2}}&\pi^+&K^+&\bar D^0\\ \pi^-&\frac{\eta}{\sqrt{3}}+\frac{\eta^\prime}{\sqrt{6}}-\frac{\pi^0}{\sqrt{2}}&K^0&D^-\\K^-&\bar K^0&-\frac{\eta}{\sqrt{3}}+\sqrt{\frac{2}{3}}\eta^\prime&D^-_s\\D^0&D^+&D_s^+&\eta_c\end{array}\right),\\ \nonumber
  V_\mu&=\left(\begin{array}{cccc}\frac{\omega}{\sqrt{2}}+\frac{\rho^0}{\sqrt{2}}&\rho^+&K^{*+}&\bar D^{*0}\\ \rho^-&\frac{\omega}{\sqrt{2}}-\frac{\rho^0}{\sqrt{2}}&K^{*0}&D^{*-}\\K^{*-}&\bar K^{*0}&\phi&D^{*-}_s\\D^{*0}&D^{*+}&D_s^{*+}&J/\psi\end{array}\right).
 \end{align}
The channels considered are $ \{ | D \bar D; I=0  \rangle, | D^+_s D^-_s  \rangle \} $. With the isospin phase convention of the $ D$ and $\bar{D}$ doublets being $ D \equiv (D^+, - D^0)$ and $\bar{D}  \equiv  ( \bar{D}{}^0, D^-)$, then the $ | D \bar D; I=0  \rangle $ state is given by 
\begin{eqnarray}
|D \bar{D}, I=0 \rangle = \frac{1}{\sqrt{2}} \left( | D^{+}D^{-}  \rangle + | D^{0}\bar{D}^{0}  \rangle \right).             
\label{eq6}
\end{eqnarray}

We construct the meson-meson scattering amplitude using the Lagrangian (\ref{LPPV}) by exchanging vector mesons between the pseudoscalars. We neglect the dependence on the squared-momentum $q^2$ in the propagator $(q^2 - M_V^2)^{-1}$ of the exchanged vector, thus arriving to a contact term. Finally, the $S$-wave projected amplitude for the process $D \bar D (D_{s} \bar D_{s}) \to D \bar D (D_{s} \bar D_{s})$ is written as 
\begin{align}
 V_{ij}=- C_{ij} \frac{g^2}{2}\left[ 3 s - \left( m_1^2 + m_2^2 + m_3^2 + m_4^2 \right) -\frac{1}{s} (m_1^2 - m_2^2 )(m_3^2 - m_4^2 ) \right] ,\label{tamp}
\end{align}
where $m_{1(3)},m_{2(4)}$ are the masses of the incoming (outgoing) particles in the channel, and the coefficients $C_{ij}$ ($i,j$ running over the channels) are given by the matrix,
\begin{align}
 C =\left(\begin{array}{cc} 
 \frac{1}{2}\left( \frac{3}{m_{\rho}^2} + \frac{1}{m_{\omega}^2} + \frac{2}{m_{J/\psi}^2 } \right) &  \frac{\sqrt{2}}{m_{K^{\ast}}^2} \\ 
 \frac{\sqrt{2}}{m_{K^{\ast}}^2} &  \frac{1}{m_{\phi}^2 } + \frac{1}{m_{J/\psi}^2 } 
 \end{array}
 \right).
 \label{cij}
\end{align}
The use of the simplified versions of the vector meson propagators in Eq.~(\ref{cij}) can be justified as follows (see the details in Ref.~\cite{Bayar:2022dqa}). In the diagonal terms of the matrix $C$, at sufficiently small energies $q^0$ and momenta $\bm{q}$, the exchanged vector propagator becomes $(-1/m_V^2)$. But for the non-diagonal transitions, i.e. $D \bar D \to D_s \bar D_s$, at the $D_s \bar D_s$ threshold the exchanged momentum becomes $q^2 = (p_D - p_{D_s})^2 = p_D^2 + p_{D_s}^2 - 2 p_D p_{D_s} = m_D^2 - m_{D_s}^2 $; as a consequence, the propagator in the $C_{12}$ element is $ (q^2 - m_{K^{\ast}}^2)^{-1} = (m_D^2 - m_{D_s}^2 - m_{K^{\ast}}^2)^{-1} $, giving a factor of the order of $0.67$. Interestingly, the presence of the non-diagonal term $V_{12}$ multiplied by this factor 0.67 gives rise to the bound $D_s \bar D_s$ state. Thus, we follow Ref.~\cite{Bayar:2022dqa} and include this factor in the non-diagonal terms.

The unitarized scattering amplitude matrix $T$ is then obtained by solving the Bethe-Salpeter equation,
\begin{align}
T=V+VGT,\label{BSE}
\end{align}
where $G$ is the loop function for two intermediate mesons in a given channel~\cite{Gamermann:2006nm}, which in the dimensional regularization scheme can be written as
\begin{align}\nonumber
G_k\left(\sqrt{s}\right)
&=\frac{1}{16\pi^2}\Biggl[a_k\left(\mu\right)+\ln{\frac{m_1^2}{\mu^2}}+\frac{m_2^2-m_1^2+s}{2s}\ln{\frac{m_2^2}{m_1^2}}\Biggr.\\
& + \Biggl. \frac{q}{\sqrt{s}}\Biggl(\ln{\frac{s-\left(m_1^2-m_2^2\right)+2q\sqrt{s}}{-s+\left(m_1^2-m_2^2\right)+2q\sqrt{s}}} + \ln{\frac{s+\left(m_1^2-m_2^2\right)+2q\sqrt{s}}{-s-\left(m_1^2-m_2^2\right)+2q\sqrt{s}}}\Biggr)\Biggr],\label{gfn}
\end{align}
with $q$ being the on-shell momentum of the $k$-th channel for the CM energy $\sqrt{s}$; $m_1, m_2$ standing for the mass of the mesons in the channel; $a_k(\mu)$  denoting respectively a subtraction constant needed to regularize the divergent nature of the $G$ function, usually fitted to the data; and $\mu$ the regularization scale.

Choosing the values for the regularization parameters as in Ref.~\cite{Bayar:2022dqa}: $\mu=1500$ MeV, $a_{D \bar D}(\mu)=-1$ and $a_{D_s \bar D_s}(\mu)=-1.45$, the resulting amplitudes show the presence of two poles. The first is around $3699$~MeV with a negligible width, coupled strongly to the $D \bar D$ channel and weakly to the $D_s \bar D_s$ channel, while the other one is around $(3932-i6)$~MeV, coupling more strongly $D_s \bar D_s$ channel, and can be identified to the $X(3930)$ state. 

We focus here on the first pole---which we will denote as ``lower energy pole'' to distinguish from the one associated to the $X(3930)$---which might be interpreted as the $I=0$ $D \bar D$ (quasi-)bound state $X(3700)$. It should be mentioned that Ref.~\cite{Gamermann:2006nm} has also obtained a pole around $3722$~MeV. Their formalism was based on a potential derived from an $SU(4)$ extension of the $SU(3)$ chiral Lagrangian with an explicit $SU(4)$ breaking for the terms exchanging
charm, but considering other pseudoscalar-pseudoscalar channels beyond the $D \bar D$ and $D_s \bar D_s$, namely: $\pi \pi$, $K \bar K$, $\eta \eta$, $\eta \eta^{\prime}$, $\eta^{\prime} \eta^{\prime}$,  $\eta_c \eta$ and $\eta_c \eta^{\prime}$. The effects of these channels near the $D \bar D$ threshold are not relevant and their respective couplings are suppressed. However, these channels yield a
imaginary part to the $D \bar D$ amplitude below the threshold, providing a width to the $D \bar D$ bound state of about 36 MeV. As a consequence, for the sake of simplicity, in Refs.~\cite{Dai:2015bcc,Wang:2019evy,Dai:2020yfu} all light channels included in Ref.~\cite{Gamermann:2006nm} were accounted for in only one additional channel $\eta \eta$ ($\eta$ because this is the dominant one), with the transition potential $ D \bar D \to \eta \eta$ adjusted to produce a similar width. In this sense, a similar scheme is adopted here: the channel $\eta \eta $ is added to the ones employed in the transition potentials in Eq.~(\ref{tamp}), considering the additional elements $V_{ D \bar D , \eta \eta} = v $, $V_{ D_s \bar D_s , \eta \eta} = 0 $ and $V_{\eta \eta , \eta \eta} = 0 $, where $v$ is taken as a free parameter. In the case of the regularization parameter of $G_{\eta \eta}$, we use $a_{\eta \eta}(\mu)=a_{D \bar D}(\mu)$. 

\begin{table}[ht!]
\caption{Mass and width of the lower energy pole dynamically generated, and the couplings $g_i$ to the different channels, for different values of $a_{D \bar D}$ (the subtraction constant) and $v$ (the magnitude of the transition potential $V_{ D \bar D , \eta \eta}$). }\label{tablemwg}
\begin{ruledtabular}
\begin{tabular}{cccccc}
$a_{D \bar D}(\mu)$ & $ v$ & $ (m , \Gamma) $ & $g_{D \bar D}$ & $g_{D_s \bar D_s}$ & $g_{\eta \eta}$ \\
&  &  [MeV] &   [MeV]  &  [MeV] &  [MeV] \\
\hline
$-1 $ & 0 & $(3699.1,0)$ & $-14508.2-i0.3$ & $-5697.1-i0.5$ & - \\
$-1 $ & 100  &  $(3699.5,12.6)$ & $-14503.7+i894.1$ & $-5766.3-i896.2$ & $ 1540.1+i198.5$  \\
$-1 $ & 180  &  $(3694.2,39)$ & $-15577.4+ i2349.5$ & $-4923.8-i2963.3$ & $ 2528.6+i1259.4$  \\
$-0.8 $ & 0 &  $(3725.4,0)$ & $-10065.1-i1.3$ & $-4172.2-i0.4$ & -  \\
$-0.8 $ & 100  & $(3725.9,5.2)$ & $-10011.2+i852.1$ & $-4284.2-i490.4$ & $1010.2+i102.2$  \\
$-0.8 $ & 180 &  $(3724.5,17.9)$ & $-10990.8-i2442.4$ & $-4146.5-i19140.8$ &  $1806.7+i737.9$\\
\end{tabular}
\end{ruledtabular}
\end{table}

The results obtained for the lower energy pole  with the assumptions summarized above are shown in Table~\ref{tablemwg}. 
The last three columns display the couplings, which are obtained from the residues of the $T$ matrix, i.e. 
\begin{eqnarray}
  g_ig_j = \lim_{s \to s_0}{ (s-s_0) \ T_{ij}},
  \label{gigj}
\end{eqnarray}
where $s_0$ is the square of the bound-state mass. These couplings demonstrate that this pole couples more strongly to the $D \bar D$, as expected. Also, it can be seen that the modification of the magnitude of the transition potential $V_{ D \bar D , \eta \eta} $, encoded in the adopted value of $v$, affects mostly the imaginary part and the couplings, yielding a bigger width as it increases. Besides, the change of the regularization parameter $a_{D \bar D}(\mu)$ also influences strongly on the position of the pole and therefore on the value of the binding energy. As a consequence, noticing that the $X(3700)$ is not yet established, the variation of these two parameters will allow us to prospect how the relevant properties and observables characterizing this state might be affected by the different features of the pole associated with it.

\begin{figure}[h!]
    \centering
    \includegraphics[width=0.32\textwidth]{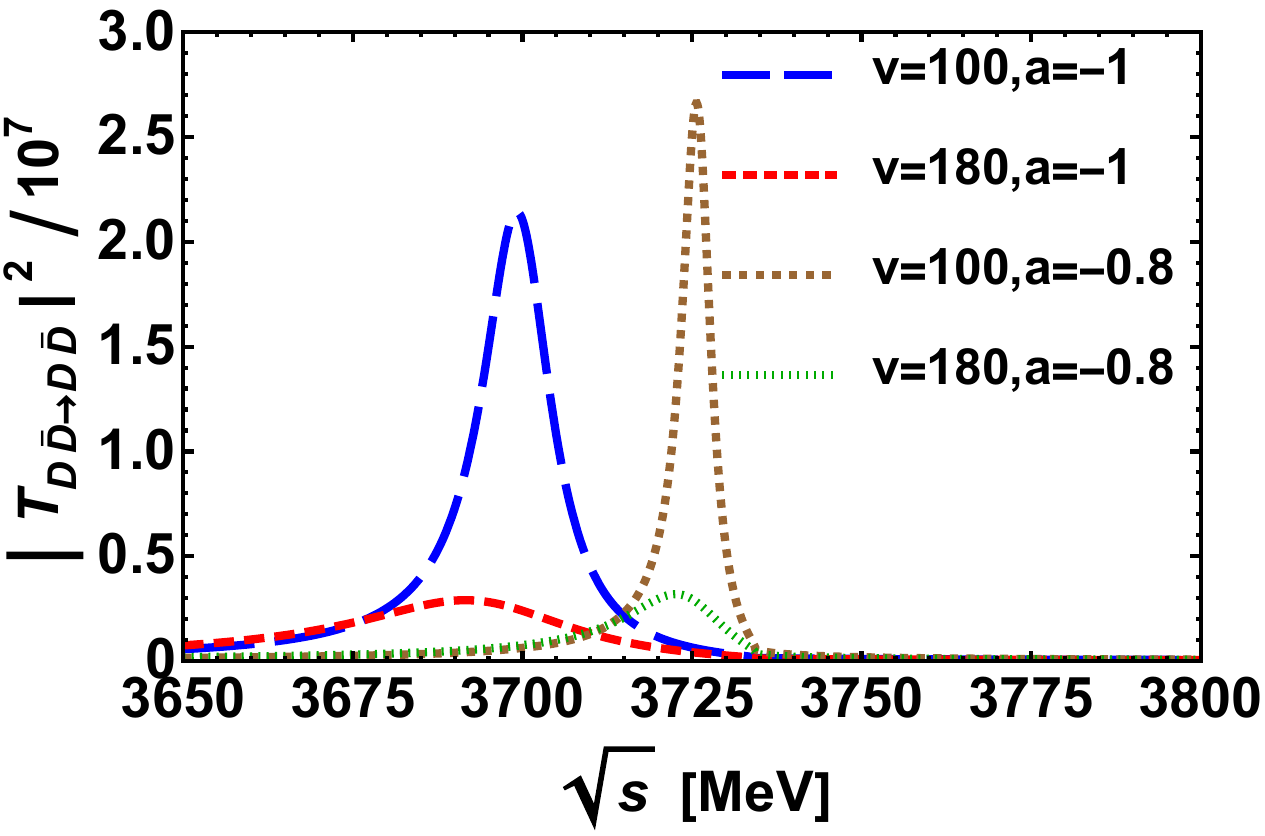}~\includegraphics[width=0.32\textwidth]{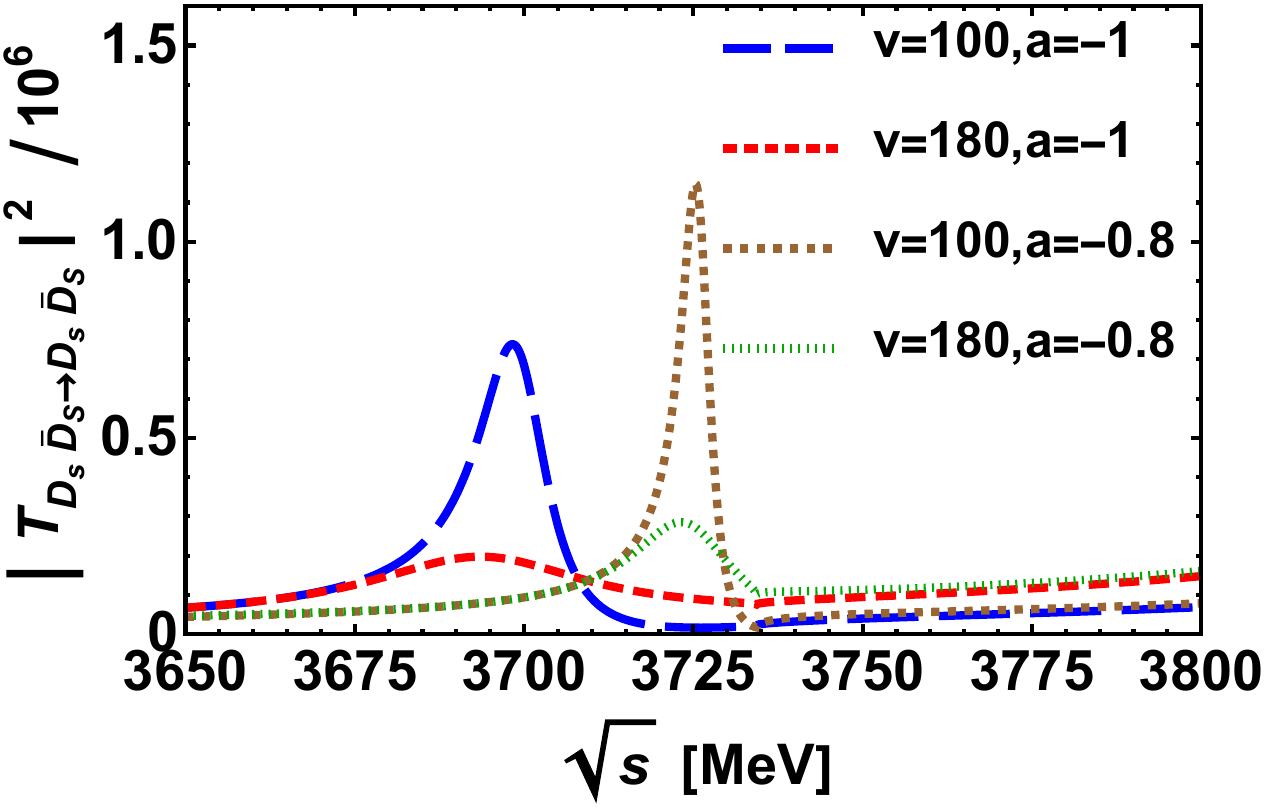}~\includegraphics[width=0.32\textwidth]{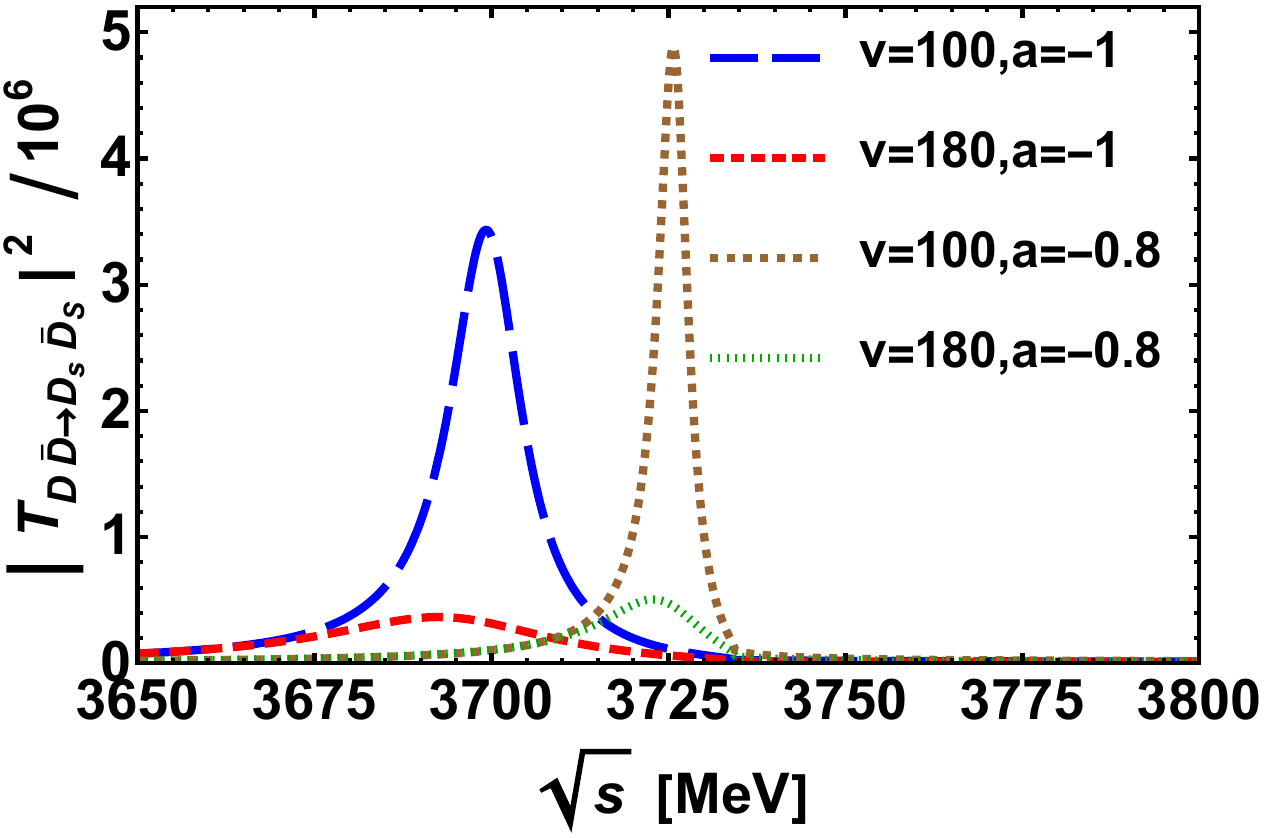}
    \caption{The squared amplitudes involving the $D \bar D$ and $D_s \bar D_s$ channels as a function of the CM energy. From left to right:  $D\bar{D} \rightarrow D\bar{D},D_s\bar{D}_s \rightarrow D_s\bar{D}_s$ and $D\bar{D} \rightarrow D_s\bar{D}_s$. }
    \label{poleRealaxes}
\end{figure}

For completeness, Fig.~\ref{poleRealaxes} shows the squared amplitudes involving the $D \bar D$ and $D_s \bar D_s$ channels representing the projection of the lower energy pole presented in Table~\ref{tablemwg} on the real axis. Besides the fact that the elastic channel $D \bar D \to D \bar D$ gives the most important contribution, the main aspects of the behavior of the pole with the relevant parameters are evidenced: the location of the peak associated to the low-energy pole is shifted to the right and becomes more prominent as $a_{D \bar D}(\mu)$ increases, whereas the growth of $v$ spreads it. 

We end this section with the discussion of another relevant observable and other quantities. The observable is the scattering length, which can be calculated for different channels through the relation~\cite{Dai:2023cyo,Feijoo:2023sfe,Khemchandani:2023xup}
\begin{align}
  a_{0}(i)=\left. \frac{T_{ii}}{8\pi \sqrt{s}} \right|_{s=s_{\textrm{thr}}} ,   \label{alength}
\end{align}
where $s_{\textrm{thr}}$ is the squared CM energy of the channel at the threshold. 
Besides, once the couplings are obtained through the use of Eq.~(\ref{gigj}), then following Refs.~\cite{Dai:2023cyo,Feijoo:2023sfe} the molecular probabilities for the channels can also be calculated as,
\begin{align}
  P_{i}=-g_i^2 \left. \frac{\partial G_{i}}{\partial s } \right|_{s=s_0}.  
  \label{prob}
\end{align}
Consequently, the nonmolecular part (that might also estimate possible missing contributions of  coupled channels) can be estimated by means of the relation  $ Z= 1 - \sum_i P_{i}. $

Table~\ref{scatprob} displays the scattering lengths $a_{0}(i)$ of the lower energy pole and the probabilities $P_i$ for the different channels, taking different values for $a_{D \bar D}$ and $v$. It can be noted that in the cases with $v \neq 0$, $a_{0}(D \bar D)$ acquires a complex contribution because the channel $ D \bar D \to  \eta \eta$ is open at the $ D \bar D $ threshold. Also, the magnitudes of $a_{0}(D \bar D)$ and $a_{0}(D_s \bar D_s)$  increase with increasing $v$ and the decrease of the regularization parameter $a_{D \bar D}$. In particular, these findings mean that as the pole approaches the $ D \bar D $ threshold, the scattering lengths $a_{0}(D \bar D)$ and $a_{0}(D_s \bar D_s)$  tend to grow. In other words, a scattering length  $a_{0}(D \bar D)$ characterizing a shallow $ D \bar D (I=0)$ bound state with smaller binding energy is larger than the typical length scale of $1 \ \mathrm{fm}$ for strong interactions. This is similar to what happens with the $ D \bar D^* $ and $ D  D^* $ interactions for the $X(3872)$ and $T_{cc}^+(3875)$ states~\cite{Dai:2023cyo,Kamiya:2022thy}. Besides, the values of the probabilities $P_1, P_2 $ and $P_3$ indicate the clear molecular state dominantly made of the  $ D \bar D (I=0)$ component, corroborating with the result coming from the couplings. 


\begin{table}[ht!]
\caption{Scattering lengths $a_{0}(i)$ of the lower energy pole dynamically generated and the probabilities  $P_i$ for the different channels, taking different values for $a_{D \bar D}$ (the regularization parameter) and $v$ (magnitude of the transition potential $V_{ D \bar D , \eta \eta}$). }\label{scatprob}
\begin{ruledtabular}
\begin{tabular}{ccccccccc}
$ (a_{D \bar D}(\mu), v) $ & $ (m , \Gamma) $ & $a_{0}(D \bar D)$ & $a_{0}(D_s \bar D_s)$ & $a_{0}(\eta \eta )$ & $P_1$ &  $P_2$ &  $P_3$ \\
  &  [MeV] &   [fm]  &  [fm] & [fm] & & &  \\
\hline
$(-1, 0)$ & $(3699.1,-)$ & $0.81$ & $1.42+i0.62$ & $-$ & 0.93 & 0.04  & $-$   \\
$(-1,100)$  &  $(3699.5,12.6)$ & $0.80+i0.06$ & $0.86+i0.61$ & $-0.25$ & 0.93 & 0.04 & $0.001$  \\
$(-1,180)$  &  $(3694.2,39)$ & $0.73+i0.13$ & $0.62+i0.31$ & $ 1.66$ & 0.97 & 0.02 & $0.002$ \\
$(-0.8,0)$  &  $(3725.4,-)$  & $1.57$ & $1.34+i 0.54$ & $-$ & 0.96 & 0.02 & $-$  \\
$(-0.8,100)$  & $(3725.9,5.2)$ & $1.56+i0.22$ & $0.86+i0.58$ & $-0.23$ & 0.96 & 0.03 & 0.001  \\
$(-0.8,180)$  & $(3724.5,17.9)$ & $1.23+i0.43$ & $0.61+i0.31$ &  $2.76$ & 0.97 & 0.02 & 0.001 \\
\end{tabular}
\end{ruledtabular}
\end{table}

Next, we discuss how these amplitudes and scenarios manifest in the femtoscopy correlation functions.

\section{Correlation Function}\label{sec-cf}


The correlation function (CF) constitutes the fundamental observable of the study of femtoscopic correlations. For two-particle systems, the CF is characterized as the ratio of the probability of measuring the two-particle state and the product of the probabilities of measuring each individual particle~\cite{Fabbietti:2020bfg}. From the theoretical side, the correlation function will be accessed from the Koonin–Pratt formula~\cite{Lisa:2005dd,Koonin:1977fh,Pratt:1986cc},
\begin{eqnarray}
C(k) & = & \int d^3 r \ S_{12}(\bm{r}) \ \vert  \Psi (\bm{k} ; \bm{r}) \vert ^2, 
\label{cf1}
\end{eqnarray}
where $ \bm{k} $ is the relative momentum in the CM frame of the pair; $\bm{r}$ is the relative distance between the two particles; $\Psi (\bm{k} ; \bm{r})$ is the relative two-particle wave function; and $ S_{12}(\bm{r}) $ is the source function. 

In order to relate the CF in~(\ref{cf1}) to the coupled-channel formalism described in the previous section, it is more convenient to make use of the Koonin–Pratt formula generalized to coupled channels~\cite{Vidana:2023olz,Feijoo:2023sfe,Albaladejo:2023pzq,Khemchandani:2023xup,Abreu:2024qqo}, i.e.
\begin{eqnarray}
C_i(k) & = & 1 + 4 \pi \theta (q_{\textrm{max}} - k) \int_{0}^{\infty} d r r^2 S_{12}(\bm{r}) \left( \sum_j w_j \vert j_0(kr) \delta_{ji} + T_{ji}(\sqrt{s}) \widetilde{G}_j(r; s) \vert^2 - j_0^2(kr) \right) ,  \nonumber \\
\label{cf2}
\end{eqnarray}
where  $i$ denotes a specific channel, $ w_j $ is the weight of the observed channel $ j $; $ j_{\nu}(kr) $ is the spherical Bessel function; $\sqrt{s}$ is the CM energy; $k = \lambda^{1/2} (s,m_{1}^2,m_{2}^2)/(2\sqrt{s})$ is the relative momentum of the channel $i$, with $\lambda$ being the K\"allen function and $m_{1}, m_{2}$ the masses of the mesons in channel $i$; $ T_{ji} $ are the on-shell elements of the $l=0$ scattering matrix for the meson–meson interaction discussed previously; and the $\widetilde{G}_j(r; s)$ functions read
\begin{eqnarray}
\widetilde{G}_j(r; s) & = & \int\limits_{\vert \bm{q} \vert < q_{\textrm{max}} } \frac{d^3 q}{(2\pi)^3} \frac{\omega_{1j} + \omega_{2j} }{2 \omega_{1j} \omega_{2j} } \frac{j_0(qr)}{s - \left( \omega_{1j} + \omega_{2j} \right)^2 +i \varepsilon} ,   
\label{gtilde}
\end{eqnarray}
with $ \omega_{aj} \equiv \omega_{aj}(q) = \sqrt{q^2 + m_a^2}$ being the energy of the  $a$-th particle in the  $j$-th channel, and $q_{\textrm{max}}$ being a sharp cutoff momentum introduced to regularize the $r \to 0$ behavior. As in the precedent works~\cite{Khemchandani:2023xup,Abreu:2024qqo}, $q_{\textrm{max}}$ is chosen to be $q_{\textrm{max}} = 700 \  \mathrm{MeV}$, but it should be noticed that $j_0(qr)$ suppresses the dependence of the integrand with large $q$. 

Concerning the $r$-dependence for the source function, the usual static Gaussian profile normalized to unity is employed~\cite{Lisa:2005dd,ALICE:2021cpv}, i.e.
\begin{eqnarray}
S_{12}(\bm{r})  & = & \frac{1}{\left(  4 \pi \right)^{\frac{3}{2}} R^3} \exp{\left(  -\frac{r^2}{4 R^2 }\right)}. 
\label{sourcef1}
\end{eqnarray}
The source size parameter $R$ commonly spans from 1 fm (high-multiplicity events in pp collisions) to 5 fm (central PbPb collisions). Other choices of source functions can be found in Refs.~\cite{Kuroki_2024,Lisa:2005dd}, but here we will only use the single Gaussian form of Eq.~\eqref{sourcef1}.

To obtain experimentally accessible CFs, the $D \bar D (I=0)$ system should be expressed in terms of the fixed charge states, given in Eq.~(\ref{eq6}).
Therefore, two CFs for the two fixed charge states should be introduced. In the case of the neutral $ \left\vert  D^0 \bar D^0  \right\rangle $ system, only the strong contribution coming from the Bethe-Salpeter equation in Eq.~(\ref{BSE}) is relevant, but with a factor $\frac{1}{2}$ encoded in $ T_{D \bar D \to D \bar D }^{(I=0)} $ calculated in the previous section to account for the projections of the fixed charge states into the $\vert D \bar D, I=0 \rangle $ system. For the other reactions, the amplitude $ T_{D \bar D \to i }^{(I=0)} $ is multiplied by a factor $1/\sqrt{2}$. 

Importantly, for the $\left\vert D^+ D^- \right\rangle $ state, the Coulomb interaction must be taken into account. In that regard, it is worth mentioning here the analysis of the meson-baryon CFs done in Ref.~\cite{Encarnacion:2024jge}. The Coulomb contribution depends on the momenta of the mesons in the incoming and outgoing channels, so it cannot be factorized on-shell out of the integral over momenta in the complete version of the Bethe-Salpeter equation (\ref{BSE}). Notwithstanding, some checks have been made to show that is a reasonable approximation to proceed as follows: the integral for the first-order Coulomb amplitude is explicitly calculated, and the strong amplitude is factorized on-shell out of the integral over momenta and unitarized. 
Besides, Ref.~\cite{Kamiya:2022thy} analyzed the nature of the $T_{cc}^+(3875)$ and $X(3872)$ states considering two separate CFs for the strong and Coulomb contributions. Its findings sound plausible, despite the possible difference related to the wave function due to the separate calculations. 
Thus, in view of the exploratory nature of this study, the two CFs for the two fixed charge states are given by
\begin{eqnarray}
C_{ D^0 \bar D^0}(k) & = & C_{ D^0 \bar D^0}^{\textrm{(S)}}(k) \ , \nonumber  \\
C_{ D^+ D^-}(k) & = & C_{ D^+ D^-}^{\textrm{(S)}}(k) + C_{ D^+ D^-}^{\textrm{(C)}}(k) -1\ ,
\label{cf3}
\end{eqnarray}
where $C_{ i }^{\textrm{(S)}}(k)$ is the pure strong contribution calculated with the amplitudes previously introduced, and $C_{ D^+ D^-}^{\textrm{(C)}}(k)$ is the Coulomb contribution. 
Since the pure Coulomb interaction is factorized in the correlation function, the $C^{\textrm{(C)}}_{D^+D^-}(k)$ component can be calculated using the complete Coulomb wave function~\cite{joachain1975quantum} (where we use the same notation as in~\cite{Torres-Rincon:2023qll}),
\begin{equation}
    \Phi^{\textrm{C}}(r,z;k)= e^{-\pi \gamma/2} \Gamma(1+ i\gamma) e^{ikz} \ {_1}F_{1}(-i\gamma;1;ik(r-z)) \ , \label{eq:completeCoulomb}
\end{equation}
into the Koonin-Pratt equation. In Eq.~\eqref{eq:completeCoulomb}, $\Gamma(z)$ is the Euler gamma function, ${_1}F_{1}(x,y;z)$ the confluent hypergeometrical function or Kummers function, $\gamma$ is the Sommerfeld factor,
\begin{equation} 
\gamma=Z_1 Z_2 \frac{\mu \alpha}{k} \label{eq:sommerfeld}
\end{equation}
with $Z_1Z_2$ the product of the charges, $\alpha$ the fine structure constant, and $\mu$ is the reduced mass of the hadron pair.

The last ingredients in Eq.~(\ref{cf3}) to be remarked are the weights $w_j$'s. They are related to the multiplicity of the pairs yields from primary particles generated in the collision. In the present situation, since the relevant channels have charmed mesons, the simplest choice is adopted with $w_j=1$, with $j=D^0 \bar{D}^0, D^+ D^-, D_s^+ D_s^-, \eta \eta$. In principle, we expect $w_{\eta \eta}$ should acquire a higher value, due to the higher degeneracy of the light states, but one must notice that the wave function connecting this channel to the observed one will provide a small contribution to the correlation function since the coupling is rather small compared to the direct channel, as evidenced by the tiny molecular probability $P_3 \ll P_1,P_2$. Therefore, we expect that the contribution of a $w_{\eta \eta}$ not equal to one does not play a relevant role in the CF.

\begin{figure}[h!]
    \centering
    \includegraphics[width=0.45\textwidth]{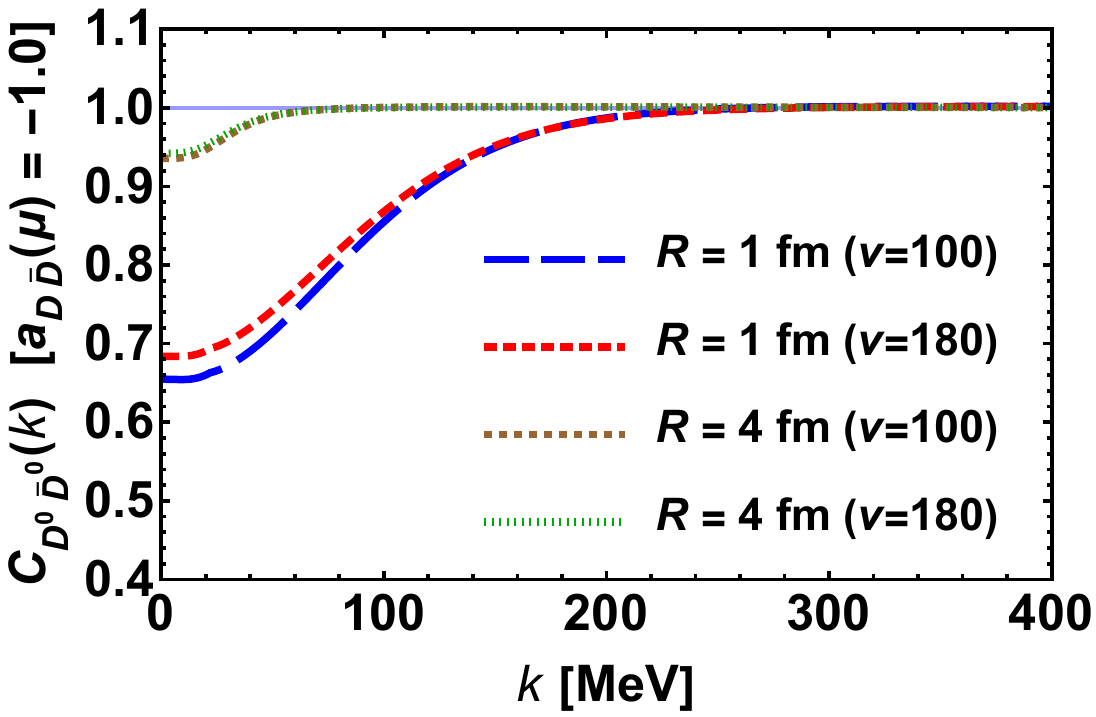}~\includegraphics[width=0.45\textwidth]{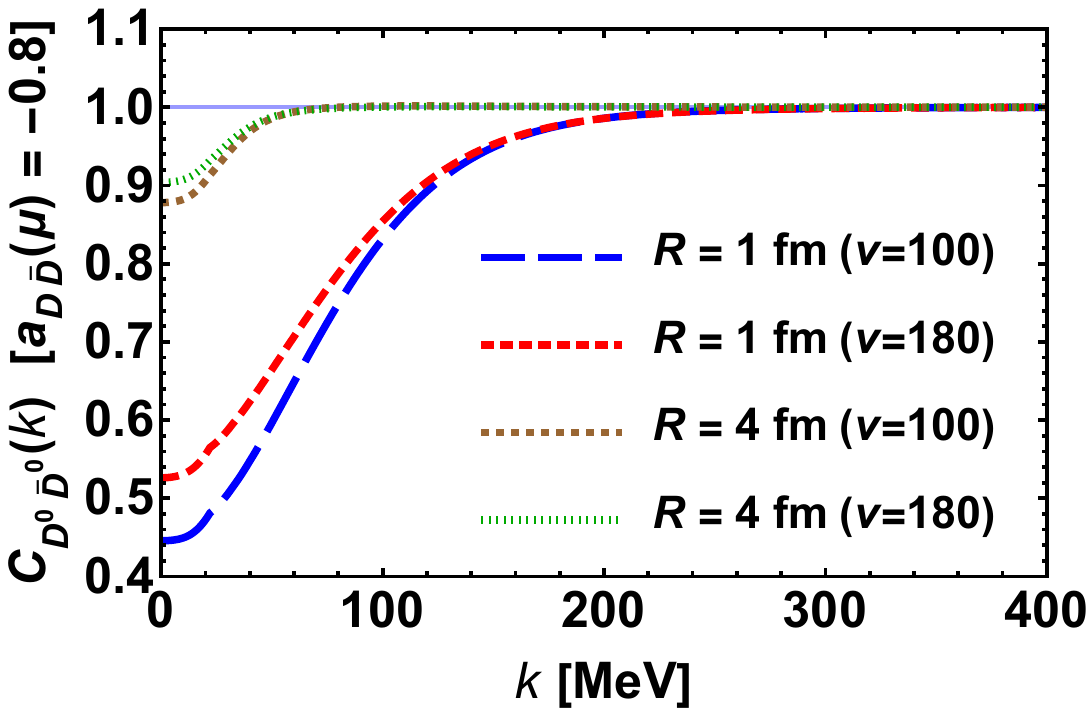}
    \caption{The $ D^0 \bar D^0$ CF defined in Eq.~(\ref{cf3}) as a function of the CM relative momentum $k$, taking different values of the parameters $R, v$ and $a_{D \bar D}(\mu)$. }
    \label{fig:cf1}
\end{figure}

After introducing the necessary elements to calculate the CFs, now the results can be presented and analyzed. In Fig.~\ref{fig:cf1} the $ D^0 \bar D^0$ CF defined in Eq.~(\ref{cf3}) is plotted as a function of the CM relative momentum $k$, taking different values of the parameters $R, v$ and $a_{D \bar D}(\mu)$. First, one can notice that in the low-momentum limit $C_{ D^0 \bar D^0}(k) $ presents a dip and increases as the source size parameter grows. This fact might be interpreted as the signature of a bound state, in light of the discussion summarized in Ref.~\cite{Khemchandani:2023xup} emphasizing the role of the scattering length to source size ratio~\footnote{The dependence of the correlation function for $k \rightarrow 0$ on the size of the source, for a potential with a bound state, has also been elaborated in~\cite{Mihaylov:2021glj,Fabbietti:2020bfg}.}. In the present case of an attractive interaction, from Table~\ref{scatprob} it can be seen that for smaller values of $R$ we have the ratio $ a_{0}(D \bar D) / R   \sim 1 $, at which the CF shows a stronger dip in the region near $k=0$. But larger sources imply $a_{0}(D \bar D) \ll R $, then this dip is weakened and $  C_{ D^0 \bar D^0} (0) \lesssim 1 $.  As a consequence, experimental analyses of the CF in systems with different sizes, for instance $pp, pA$, and $AA$ collisions, will help us to elucidate if this interpretation is compatible with the experimental data. 

In addition, it is also of importance to verify in Fig.~\ref{fig:cf1} that the change on the parameter $v$, which impacts mostly on the imaginary part (and slightly on the real part), yields a mild modification in the low-momentum region $C_{ D^0 \bar D^0}(k) $. Therefore, a bound state with larger width engenders a weaker dip in the CF. On the other hand, the dependence on the regularization parameter $a_{D \bar D}(\mu)$, which controls the pole position, appears to be stronger. The closer the pole to the $D \bar D$ threshold, the more intense the correlations, evidenced by more prominent dips. But one should note that if the state becomes sufficiently shallowly bound so that its scattering length is noticeably higher than the source parameter $R$, the CF would show an enhancement in the low-momentum region, similarly to the poles associated to the $X(3872)$ and $T_{cc}^+(3875)$~\cite{Kamiya:2022thy,Vidana:2023olz}.

\begin{figure}[h!]
    \centering
    \includegraphics[width=0.45\textwidth]{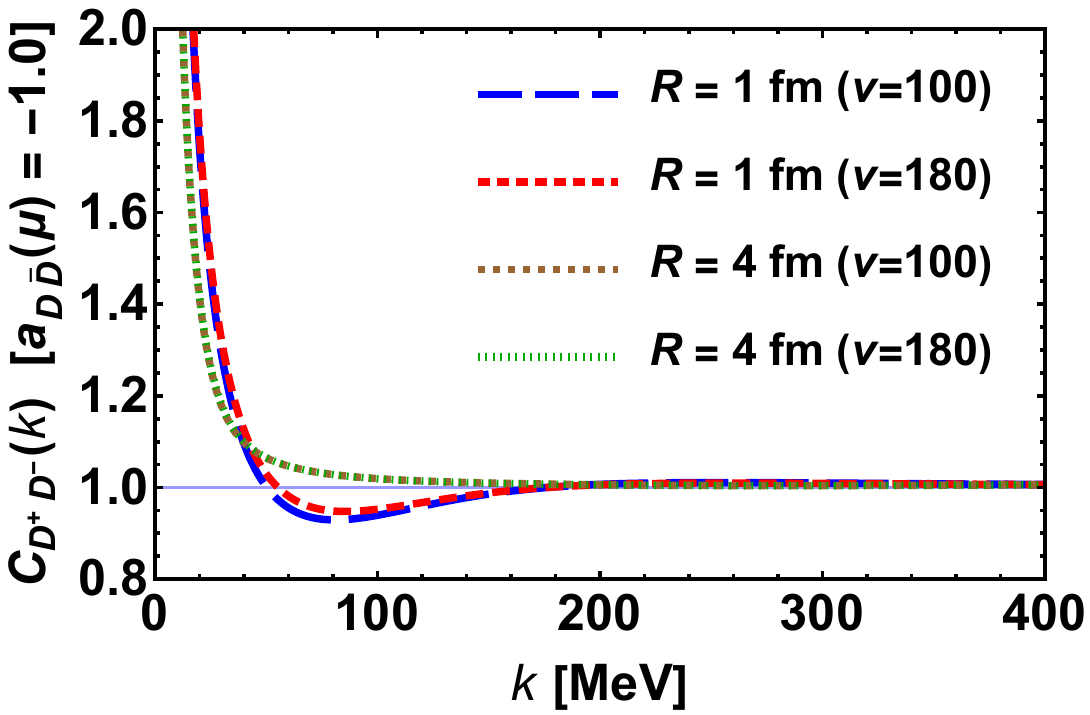}~\includegraphics[width=0.45\textwidth]{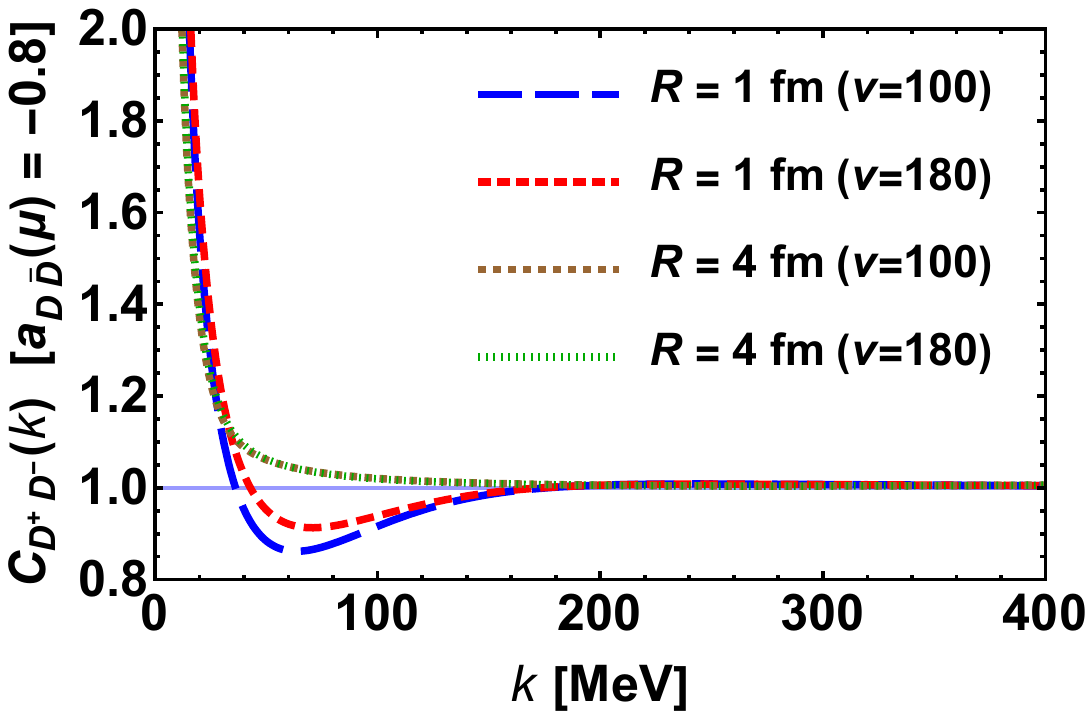}
    \caption{The $ D^+ D^-$ CF defined in Eq.~(\ref{cf3}) as a function of the CM relative momentum $k$, taking different values of the parameters $R, v$ and $a_{D \bar D}(\mu)$. }
    \label{fig:cf2}
\end{figure}

Next, we show in Fig.~\ref{fig:cf2} the $D^+ D^-$ CF defined in Eq.~(\ref{cf3}) as a function of the CM relative momentum $k$, taking different values of the parameters $R, v$ and $a_{D \bar D}(\mu)$. At small relative momentum, the attractive Coulomb interaction yields a sizable enhancement and is the dominant contribution. 
The effect of the strong interaction appears only at moderate values of $k$ by means of a dip. But this dip is weakened for deeper bound structures that are represented by smaller $a_{D \bar D}(\mu)$. On the other hand, this dip is more pronounced for shallow bound states (engendered by higher $a_{D \bar D}(\mu)$) with smaller widths (yielded by smaller $v$). Also, the contribution of the strong force is practically suppressed for large sources. Hence, our findings suggest that the $D^+ D^-$ CF is more sensitive to the signature of the $X(3700)$ if it is a narrow and weakly bound structure, produced in a smaller source environment.

\begin{figure}[h!]
    \centering
    \includegraphics[width=0.45\textwidth]{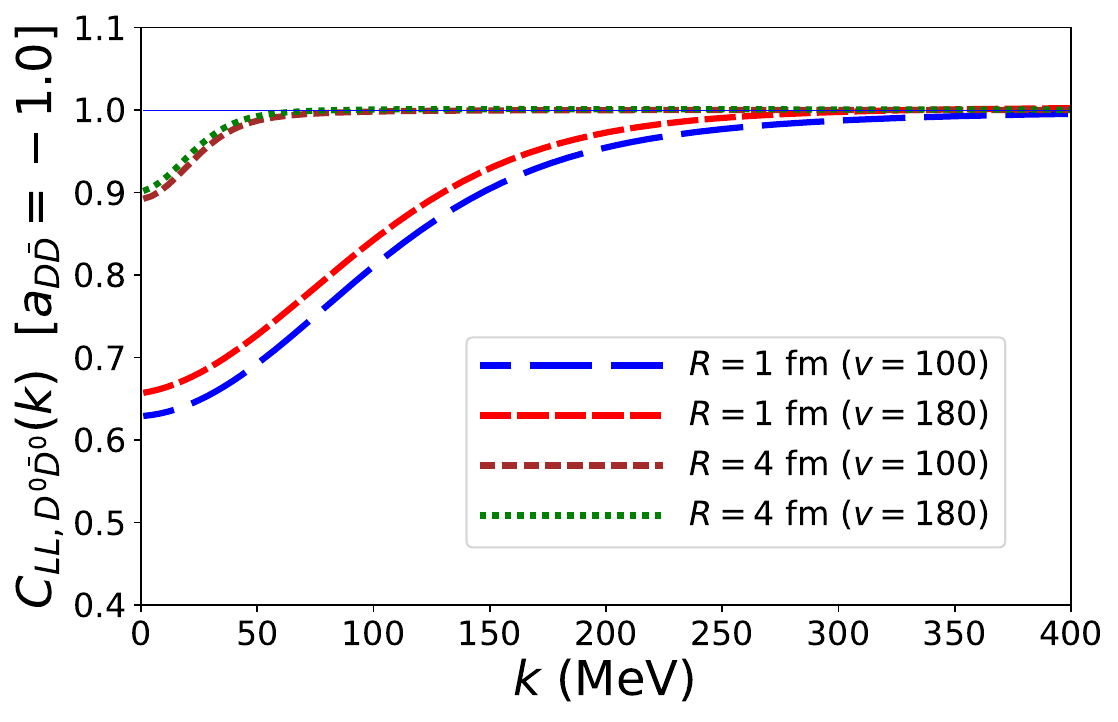}~\includegraphics[width=0.45\textwidth]{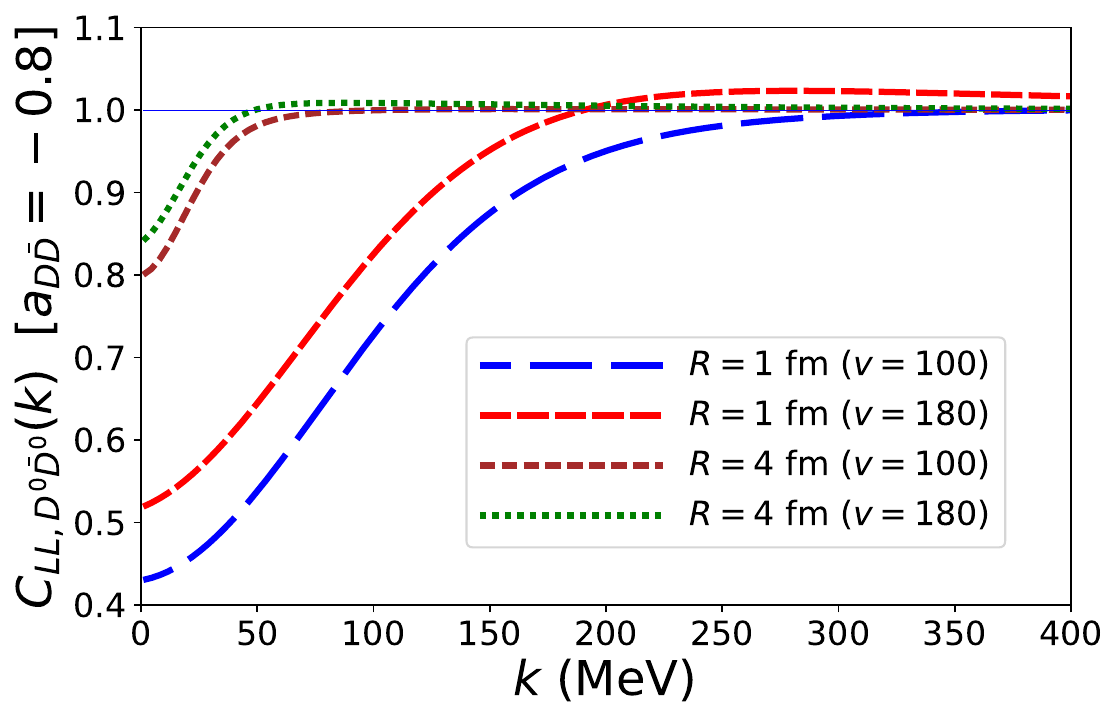}
    \caption{The $ D^0 \bar D^0$ CF in the LL approximation, according to Eq.~\eqref{eq:CkLL},for the same parameters $R, v$ and $a_{D \bar D}(\mu)$ as those shown in Fig.~\ref{fig:cf1}.\label{fig:cf1LL}}
\end{figure}

\begin{figure}[h!]
    \centering
    \includegraphics[width=0.45\textwidth]{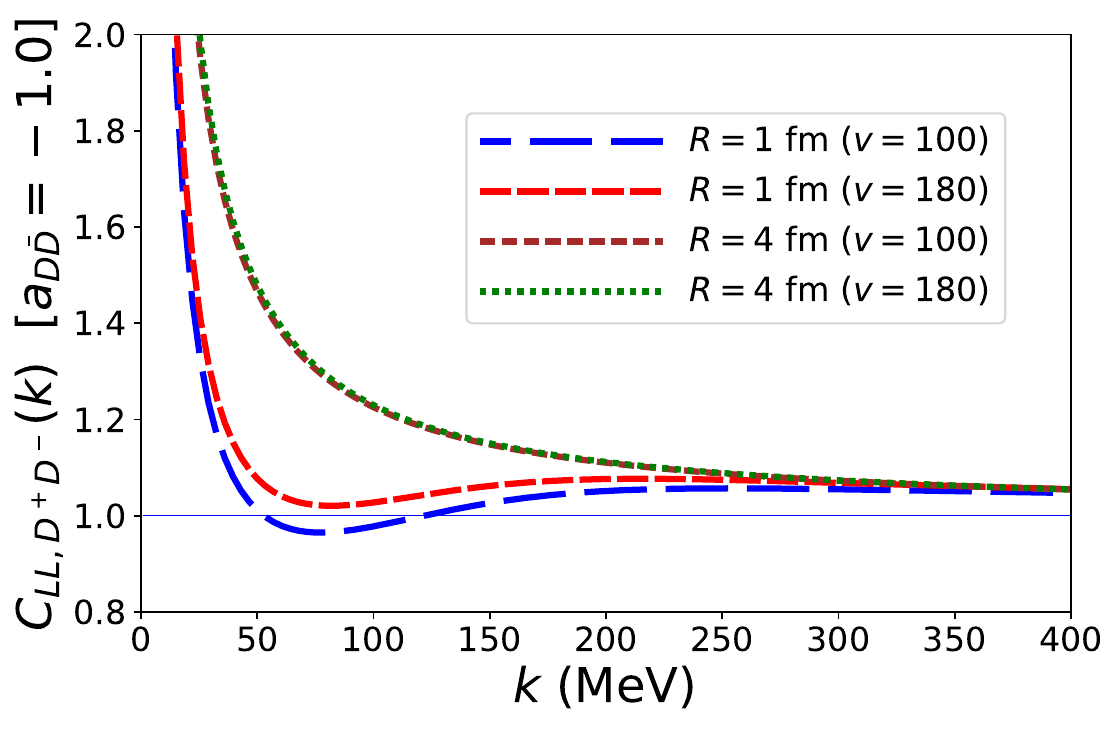}~\includegraphics[width=0.45\textwidth]{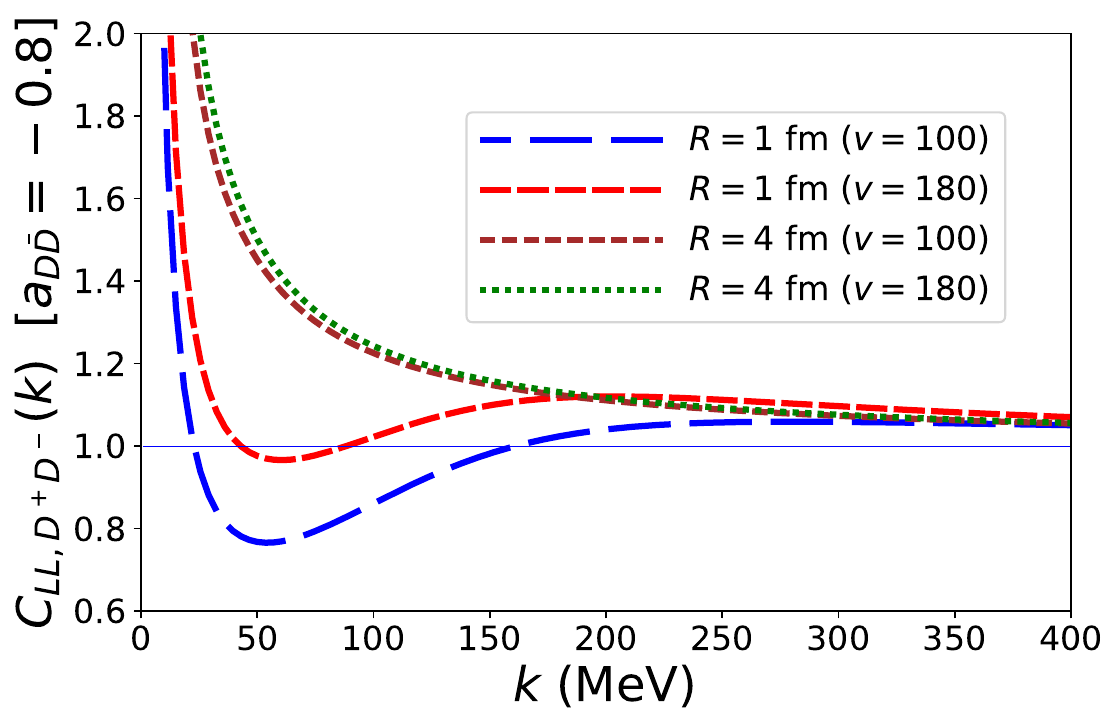}
    \caption{The $ D^+ D^-$ CF in the LL approximation, according to Eq.~\eqref{eq:CkLLCou}, for the same parameters $R, v$ and $a_{D \bar D}(\mu)$ as those shown in Fig.~\ref{fig:cf2}.\label{fig:cf2LL}}
    \
\end{figure}

For completeness, in Figs.~\ref{fig:cf1LL} and~\ref{fig:cf2LL} we present the correlation functions for the same systems given in Figs.~\ref{fig:cf1} and \ref{fig:cf2}, but using the Lednick{\'y}-Lyuboshitz (LL) formalism~\cite{Gmitro:1986ay,Lednicky:1998}. In Appendix~\ref{app:LL} we give a brief summary of this framework. 
In our results we use the zero effective range approximation.
We observe qualitative agreement with Figs.~\ref{fig:cf1} and~\ref{fig:cf2}, but with some numerical discrepancies. The most prominent difference for neutral channels is that for large sources ($R=4$ fm), the LL approximation for $a_{D \bar{D}}=-0.8$ gives a small extra correlation at small momentum ($\sim$0.8 in LL versus $\sim$0.9 in the full $T$-matrix calculation at $k=0$). For the charged channel, we observe more differences between the $T$-matrix calculation and the LL formalism. Notice that in the LL case with Coulomb interaction the correlation function goes very slowly towards $C(k)=1$ at high momentum. We have numerically checked that this is the correct asymptotic limit, but it needs very large values of the momentum. We also note that the minimum of the correlation function is more pronounced in the LL case than in the $T$-matrix calculation. Since the latter uses the full energy dependence of the interaction plus coupled channels, it should be considered as a more complete result. 

Correlation data collected by precise future measurements will be fundamental to check the hypotheses raised here and if they are compatible with a $ D \bar D (I=0)$ bound state. If so, it will enable us to determine its properties like the pole position, together with its scattering lengths and compositeness.


\section{Conclusions}

In this work, the femtoscopic $D \bar D $ correlation functions have been studied to predict the signature of the $ D \bar D $ $[X(3700)]$  bound state in the isoscalar channel. This structure has been dynamically generated by solving the coupled-channel Bethe-Salpeter equations with the local hidden-gauge formalism. Noticing that the $X(3700)$ is not yet established, we have examined how the relevant properties and observables characterizing this state---like the pole position, scattering lengths and compositeness---might be affected by the variation of the relevant parameters of the model. In particular, as the pole approaches the $ D \bar D $ threshold, the scattering lengths grow and become larger than the typical length scale of $1 \ \mathrm{fm}$ for the strong interactions, similarly to the $ D \bar D^* $ and $ D  D^* $ interactions yielding the $X(3872)$ and $T_{cc}^+(3875)$ states. 

Next, the amplitudes encoding this bound state have been employed into the momentum CFs of the $D^0 \bar D^0$ and  $D^+ D^- $ pairs. In the low-momentum limit, the $C_{ D^0 \bar D^0}(k) $ presents a dip and increases as the source size parameter grows, which can be viewed as the signature of a bound state. A pole close to the $D \bar D$ threshold will yield more intense correlations and more prominent dips, but for shallow bound states the CF would show an enhancement in the low-momentum region. In the case of the $D^+ D^-$ CF, at small relative momentum, the attractive Coulomb interaction is the dominant contribution. The strong interaction contribution appears as a dip at moderate values of $k$, which is more pronounced for shallow bound states. As a consequence, the $D^+ D^-$ CF is more sensitive to the sign of the $X(3700)$ if it is a narrow and weakly bound structure, yielded in a smaller source environment. We expect that future measurements of the $ D \bar D $ correlations for systems with different sizes will shed some light on the existence of the $X(3700)$ and its nature.

Finally, we can also provide a brief note on the intrinsic nature of the $X(3700)$. As mentioned in the Introduction, there have been intense discussions in the literature about the underlying structure of exotic states, like the $X(3872)$, which in principle could have molecular, compact tetraquark, $c \bar c $ and other components. Thus, one can wonder if these possibilities might (or might not) also be realized to the $X(3700)$ studied here. However, we stress that our focus has been on the femtoscopic CF, and therefore the hadronic interactions of a $D \bar D$ system. Accordingly, we consider a pair of heavy mesons, produced at freeze-out in heavy-ion collisions, which can interact until they (or better said, their decay products) are detected by the experiment. We find that these mesons interact dominated by a bound state below their threshold, which we interpret as the $X(3700)$. Therefore, our only claim is that this state needs to have a nonzero molecular component that can give an imprint on the CF of the $D \bar D$ pair. The relevance of other nonmolecular components will require additional and detailed studies, engendering a very distinct approach with respect to the present analysis. In that regard, it is our plan to treat in future whether this state has additional components in the Fock state.

\section{Acknowledgements}

We would like to thank Albert Feijoo and Eulogio Oset for fruitful discussions. This work was partially supported by the Brazilian Brazilian CNPq (L.M.A.: Grants No. 400215/2022-5, 308299/2023-0, 402942/2024-8), and CNPq/FAPERJ under the Project INCT-F\'{\i}sica Nuclear e Aplica\c c\~oes (Contract No. 464898/2014-5). JMT-R also acknowledges funding from the project numbers CEX2019-000918-M (Unidad de Excelencia “María de Maeztu”) and PID2023-147112NB-C21, financed by the Spanish MCIN/ AEI/10.13039/501100011033/.

\appendix
\section{Correlation function in the Lednick{\'y}-Lyuboshitz approximation~\label{app:LL}} 

Following the Lednick{\'y}-Lyuboshitz (LL) formalism given in Ref.~\cite{Gmitro:1986ay,Lednicky:1998}, one introduces the asymptotic form of the pair wave function $\psi(\bm{r};\bm{k})$, 
\begin{equation} \psi (\bm{r};\bm{k} ) = e^{i \bm{k} \cdot \bm{z}} + f (k) \frac{e^{ikr}}{r} \ , \label{eq:asymp} \end{equation}
into the Koonin-Pratt formula. In Eq.~\eqref{eq:asymp},
$f (k) \equiv f(\bm{k} \rightarrow \bm{k'})|_{|\bm{k}|=|\bm{k}'|}$ stands for the strong scattering amplitude, which is taken to be up to the $L=0$ partial wave under the effective range expansion,
\begin{equation} f^{-1} (k)= -\frac{1}{a_0} + \frac12 d_0 k^2 - i k + \mathcal{O}(k^4)   \ , \label{eq:f} \end{equation}

In addition, the LL approximation does not account for coupled channels. Therefore, we will only consider the diagonal interaction contribution. Under these approximations, the correlation function reads~\cite{Lednicky:1981su},
\begin{equation} C_{\rm{LL}} (k)=   1 +\frac{|f(k)|^2}{2R^2} + 2 \textrm{Re } f(k)\frac{F_1(2kR)}{ \sqrt{\pi} R} -  \textrm{Im } f(k) \frac{F_2(2kR)}{R} \  ,\label{eq:CkLL} \end{equation}
where $F_1(x)=x^{-1}\exp(-x^2)\int_0^x \exp(y^2) dy$ and $F_2(x)=(1-\exp(-x^2))/x$.

For the case with Coulomb interaction we refer the reader to the Appendix of Ref.~\cite{Torres-Rincon:2024znb}, where an analogous formula to~\eqref{eq:CkLL} in the Coulomb case is given. We reproduce it here for convenience,
\begin{align} C_{\rm{LL, Coulomb}} (k) & = G(\gamma) \left\{ 1 + 2 \textrm{Re } f_{\rm{C}} (k) \frac{F_1(2kR)} {\sqrt{\pi} R}
 - \textrm{Im } f_{\textrm{C}} (k) G(\gamma) \frac{F_2(2kR)}{R} \right. \nonumber \\
& \left. + \frac{|f_C (k)|^2}{2R^2} \left[ 1+ \frac12 [G(\gamma)^2 -1]
(1 -  e^{-4k^2 R^2})  \right] \right\} \ ,\label{eq:CkLLCou} \end{align}
where $G(\gamma)$ is the Gamow factor,
\be G(\gamma)=\frac{2\pi \gamma}{e^{2 \pi \gamma}-1}\ee
and $\gamma$, the Sommerfeld factor, is defined in Eq.~\eqref{eq:sommerfeld}. The scattering amplitude in the presence of the Coulomb interaction is also given in the effective range expansion~\cite{Gmitro:1986ay},
\begin{equation} f_{\rm{C}}^{-1} (k)= - \frac{1}{a_0} + \frac12 d_0 k^2 - 2 k \gamma h( \gamma^{-1})- i k G(\gamma)   \ , \label{eq:fC} \end{equation}
with
\be h(x^{-1})= - \log(|x|)+ \frac12 \psi(1-ix) + \frac12 \psi(1+ix) \ , \ee
and $\psi(x)$ the digamma function.

\bibliographystyle{apsrev4-1}
\bibliography{CF-DDBar}

\end{document}